\documentclass[]{spie}  %>>> use for US letter paper
\usepackage{amsmath,amsfonts,amssymb}
\usepackage{graphicx}
\usepackage[colorlinks=true, allcolors=blue]{hyperref}

\DeclareMathOperator{\BesselM}{I}

\title{Quantum optics in the turbulent atmosphere: Fundamental issues and applications}

\author[a,b.c]{Andrii Semenov}
\author[a]{Mykyta Klen}
\author[a]{Illia Pechonkin}
\affil[a]{Quantum Optics and Quantum Information Group, Bogolyubov Institute for Theoretical Physics of the National Academy of Sciences of Ukraine, Vulytsia Metrolohichna 14b, 03143 Kyiv, Ukraine}
\affil[b]{Department of Theoretical and Mathematical Physics, Kyiv Academic University, Boulevard Vernadskogo  36, 03142  Kyiv, Ukraine}
\affil[c]{Department of Mathematics, Kyiv School of Economics, Vulytsia Mykoly Shpaka 3, 03113  Kyiv, Ukraine}

\authorinfo{Further author information: (Send correspondence to A.A.S.)\\A.A.S.: E-mail: andrii.semenov@bitp.kyiv.ua\\ M.K.: E-mail: mykyta.klen@bitp.kyiv.ua\\I.P.: pechonkin@bitp.kyiv.ua}

\begin{document} 
\maketitle

\begin{abstract}
Quantum light propagation through turbulent atmosphere has become a subject of intensive research, spanning both theoretical and experimental studies.
This interest is driven by its important applications in free-space quantum communication, remote quantum sensing, and environmental monitoring. 
At the same time, this phenomenon itself poses an intriguing fundamental problem.
A consistent theoretical description typically makes explicit assumptions about the measurement scheme at the receiver station and/or the method of quantum-information encoding. 
A common and straightforward approach encodes the information in quantum states of a quasi-monochromatic mode, representing a pulsed Gaussian beam.
Atmospheric turbulence induces random distortions of the pulse shape and, consequently, random fluctuations of the transmittance through the receiver aperture.
These fluctuations, characterized by the probability distribution of transmittance (PDT),  directly affect the quantum state of the received light.
In this paper we examine various analytical models of the PDT, validate them through numerical simulations, and assess their range of applicability.
Furthermore, we extend the analysis beyond the standard ensemble-averaging approach, recognizing that realistic experiments typically involve time averaging.
This requires a detailed examination of the underlying random process, including the study of temporal correlations and their impact on nonclassical properties of electromagnetic radiation.
\end{abstract}


\section{INTRODUCTION}

Let us consider a typical quantum-optical experiment.
A source generates a quantum state of light. 
For simplicity, we may assume it is a single-mode state, although the discussion is not restricted to this case. 
The state is transformed and/or transmitted through a communication channel, and finally analyzed by a detection system.
The outcome is a measured value of  the observable associated with the chosen measurement device.
The experiment is repeated many times, producing a set of measured values drawn from a statistical ensemble of independent and identically prepared quantum states.
As an option, the measurement device may have variable settings.
This results in several  sets of data corresponding to different observables, each linked to a specific device setting.
In the final stage, the collected data are processed to obtain the target result.
This may involve, for example, testing Bell nonlocality \cite{Brunner2014} and optical nonclassicality \cite{titulaer65,mandel86,mandel_book,vogel_book,agarwal_book,Schnabel2017,sperling2018a,sperling2018b,sperling2020}, sharing a cryptographic key \cite{gisin02,Xu2020,Pirandola2020,Renner2023},  estimating radiation parameters  \cite{Tsang2016,Rouviere2024}, and so on. 

At first glance, atmospheric turbulence may seem to fit naturally into this picture as just another type of communication channel between the transmitter (source) and the receiver.
However, such channels possess distinctive features.
The random parameters describing the channel at a given time depend on their values at previous times.
As a result, the measurement outcomes at the receiver are no longer statistically independent, and time averaging over the experimental record can differ substantially from ensemble averaging.

A consistent description of quantum-light propagation through a turbulent atmosphere requires a clear specification of the methods used for quantum-information encoding. 
In this paper, we focus on a simple yet widely employed scenario in free-space communication, where the information is encoded in the quantum state of a quasi-monochromatic mode representing a pulsed Gaussian beam. 
In this case, the precise mode shape at the receiver station is not relevant, as we are concerned with detecting the total radiation transmitted through the receiver aperture. 
The temporal (or spectral) shape of the pulse is also not crucial, except when the detection technique at the receiver is sensitive to it, as is the case for detectors affected by dead time \cite{Semenov2024} or relaxation time \cite{Uzunova2022}.
The corresponding theoretical framework encompasses a variety of scenarios, including polarization analysis~\cite{ursin07,fedrizzi09,semenov10,gumberidze16}, photocounting measurements~\cite{perina73,milonni04}, and properly arranged homodyne detection, with the local oscillator transmitted in the same spatial mode but in orthogonal polarization~\cite{usenko12,guo2017,papanastasiou2018,derkach2020b,hosseinidehaj2019,ShiyuWang2018,chai2019,Derkach2021,pirandola2021,hosseinidehaj2021,pirandola2021b,pirandola2021c,peuntinger14,Wang2018,hofmann2019,zhang2017,villasenor2021,bohmann16,hosseinidehaj15a,elser09,heim10,semenov12,heim14}.

More technically, let $\hat{a}_{\mathrm{in}}$ denote the photon annihilation operator of the input mode, and let $\hat{a}_{\mathrm{out}}(t)$ be the photon annihilation operator of the same mode after passing through the receiver aperture at time $t$. 
These operators are related through the input-output relation
	\begin{align}
	\hat{a}_{\mathrm{out}}(t) = \sqrt{\eta(t)}\hat{a}_{\mathrm{in}} + \sqrt{1-\eta(t)}\hat{c},
	\end{align}
where $\hat{c}$ is the annihilation operator of a noise mode in the vacuum state. 
In this relation, $\eta(t)$ represents the transmittance at time $t$.
For each transmitted pulse, the corresponding quantum-state input-output relation can be written as 
	\begin{align}\label{Eq:InOut_t}
	P_{\mathrm{out}}(\alpha;t) = \frac{1}{\eta(t)} P_{\mathrm{in}}\left(\frac{\alpha}{\sqrt{\eta(t)}}\right),
	\end{align}
where $P_{\mathrm{in}}(\alpha)$ and $P_{\mathrm{out}}(\alpha;t)$ are the Glauber–Sudarshan $P$ functions \cite{glauber63c,sudarshan63} of the input mode and of the output mode at time $t$, respectively. 
This transformation of the $P$ function is the standard representation of attenuation in a lossy bosonic channel \cite{mandel_book}.

 In general, since $\eta(t)$ is a non-trivial stochastic process, the $P$ functions (and thus the corresponding density operators) at different times are not statistically independent. 
 The same holds for any observable measured at the receiver station. 
 In experiments, one typically estimate time averaging of quantum observables (e.g., photon number, field quadratures, and so on) over a finite interval $t \in [0,T]$. 
 Extending this procedure to the entire quantum state, one may introduce the time-averaged $P$ function (i.e., density operator) as
	 \begin{align}\label{Eq:InOut_T}
	 	P_{\mathrm{out}}^T(\alpha) = \frac{1}{T}\int_0^T dt\frac{1}{\eta(t)} P_{\mathrm{in}}\left(\frac{\alpha}{\sqrt{\eta(t)}}\right).
	 \end{align}
 Importantly, the time-averaged $P$ function (or density operator) does not yield correct time-averaged values of observables.
 For sufficiently large $T$ [in practice, larger than the correlation time of $\eta(t)$], the process can be regarded as ergodic. 
 In this case, time averaging can be replaced by ensemble averaging, which yields the input–output relation \cite{semenov09,vasylyev12,vasylyev16,vasylyev18,klen2023},
	 \begin{align}\label{Eq:InOut}
	 	P_{\mathrm{out}}(\alpha) = \int_{0}^1 d\eta \mathcal{P}(\eta) \frac{1}{\eta} P_{\mathrm{in}}\left(\frac{\alpha}{\sqrt{\eta}}\right),
	 \end{align}
 where $\mathcal{P}(\eta)$ is the probability distribution of transmittance (PDT), which characterizes the given free-space channel.
 This $P$ function (or the corresponding density operator) can be used to calculate the correct ensemble-averaged values of observables. 
 
 Time correlations may nevertheless play an important role even in scenarios where ensemble averaging is employed \cite{Klen2024}. 
 First, one may exploit time-bin quantum-information encoding, thereby increasing the effective Hilbert-space dimension. 
 Second, one may monitor the channel transmittance with bright classical light and send a quantum state only within short time intervals when the transmittance exceeds a certain threshold.

In this paper, we review the main analytical models of the PDT \cite{vasylyev12,vasylyev16,vasylyev18,klen2023,pechonkin2025}, classifying them according to the degree of empirical assumptions involved. 
Importantly, within our approach any analytical model should be characterized by parameters that, at least in principle, can be derived analytically. 
We also address a powerful tool for the numerical evaluation of the PDT \cite{klen2023}, which is based on the sparse-spectrum model \cite{Charnotskii2013a,Charnotskii2013b,Charnotskii2020} of the phase-screen method \cite{Fleck1976,Frehlich2000,Lukin_book,Schmidt_book}. 
Finally, we discuss temporal correlations of the transmittance $\eta(t)$ and explore their implications for quantum-state transfer through free-space channels \cite{Klen2024}.

\section{CLASSICAL RADIATION}
\label{Sec:ClassRad}

A significant advantage of using the Glauber–Sudarshan $P$ representation is that the corresponding input–output relations (\ref{Eq:InOut_t}), (\ref{Eq:InOut_T}), and (\ref{Eq:InOut}) closely resemble those in classical optics. This implies that the PDT can be derived from classical fields prepared in coherent states of the considered quasi-monochromatic mode. In practice, this means that methods of classical atmospheric optics \cite{Tatarskii,Tatarskii2016,Fante1975,Fante1980,Andrews_book} can be employed to calculate the relevant characteristics of the stochastic process $\eta(t)$.

Let us first summarize the information available from classical theory on the stochastic properties of $\eta(t)$ and related concepts. 
We start with the stochastic differential equation for the beam amplitude $u(\mathbf{r};z)$, where $\mathbf{r}=(x,y)$ and $z$ are the transverse and axial propagation coordinates, respectively. 
This is the paraxial equation \cite{Fante1975},
    \begin{align}\label{Eq:paraxial}
		2ik\frac{\partial u(\textbf{r};z)}{\partial z}+ \Delta_{\textbf{r}} u(\textbf{r};z)+2k^2 \delta n(\textbf{r};z)u(\textbf{r};z)=0,
	\end{align}
where $k$ is the wave number, $\Delta_{\mathbf{r}}$ is the transverse Laplace operator, and $\delta n(\mathbf{r};z)$ is the stochastic component, i.e., random fluctuations of the refractive index caused by atmospheric turbulence.
Typical scenarios assume the preparation of a Gaussian beam at the transmitter, which imposes the boundary condition in the form
	\begin{align}\label{Eq:BoundaryConditions}
		u(\mathbf{r};0)=\sqrt{\frac{2}{\pi
		W_0^2}}\exp\Bigl[-\frac{\mathbf{r}^2}{W_0^2}{-}\frac{ik}{2F_0}\mathbf{r}^2\Bigr],
	\end{align}
where $W_0$ and $F_0$ denote the beam-spot radius and the wavefront radius at the transmitter, respectively.

For the fluctuating part of the refraction index, the second-order correlation function in the Markovian approximation can be written as
	\begin{align}\label{Eq:Correlator_n_n}
		\langle\delta n(\mathbf{r}_1;z_1)\delta n(\mathbf{r}_2;z_2)\rangle
		=\int_{\mathbb{R}^2}d^2\boldsymbol{\kappa}\Phi_n(\boldsymbol{\kappa};\kappa_z{=}0)e^{i\boldsymbol{\kappa}\cdot(\mathbf{r}_1-\mathbf{r}_2)}\delta(z_1-z_2),
	\end{align}
where $\Phi_{n}(\boldsymbol{\kappa};\kappa_z)$ is the turbulence power spectral density.
The simplest model is the Kolmogorov turbulence spectrum, which remains convenient for many analytical calculations. 
A more realistic model is the modified von K\'arm\'an--Tatarskii spectrum, given by
	\begin{align}\label{Eq:Karman}
		\Phi_n(\boldsymbol{\kappa};\kappa_z)=\frac{0.033 C_n^2\exp\left[-\left(\frac{\kappa\ell_0}{2\pi}\right)^2\right]}{\left(\kappa^2+L_0^{-2}\right)^{11/6}},
	\end{align}
where $C_n^2$ is the index-of-refraction structure constant---the key parameter characterizing turbulence strength, while $L_0$ and $\ell_0$ denote the outer and inner turbulence scales, respectively. 
Here $\kappa=\sqrt{\boldsymbol{\kappa}^2+\kappa_z^2}$. 
The Kolmogorov spectrum is recovered in the limit $L_0\to\infty$ and $\ell_0\to 0$.

We consider experimental procedures with phase-insensitive measurements or, in the case of homodyne detection, with phase-preserving settings \cite{elser09,heim10,semenov12}. Consequently, our primary interest lies in the intensity at the receiver aperture plane,
	\begin{align}
		I(\textbf{r};L)=|u(\textbf{r};L)|^2,
	\end{align}
 where $L$ is the channel length.
The transmittance efficiency is defined as the fraction of the intensity transmitted through the aperture opening $\mathcal{A}$,
	\begin{align}\label{Eq:eta}
		\eta =\int_{\mathcal{A}} d^2\textbf{r} I(\textbf{r};L).
	\end{align}
Throughout this work, we assume that the intensity $I(\mathbf{r};L)$ is normalized in the transverse plane.

\begin{figure} [ht!]
   \begin{center}
   \begin{tabular}{c} %% tabular useful for creating an array of images 
   \includegraphics[height=5cm]{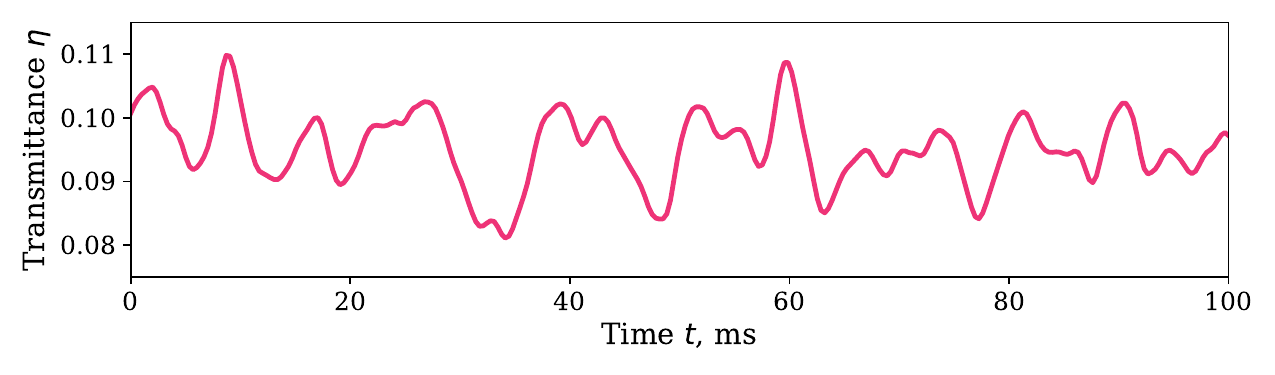}
   \end{tabular}
   \end{center}
   \caption[example] 
%>>>> use \label inside caption to get Fig. number with \ref{}
   { \label{Fig:StochasticProcess} 
Typical behavior of $\eta(t)$ obtained from numerical simulations. Constant losses due to absorption, scattering, and imperfections of the optical system are not included.}
\end{figure}

Clearly, random fluctuations of the refractive index $\delta n(\mathbf{r};z)$ lead to random variations of the transmittance $\eta$.
It is therefore essential to understand how these fluctuations of $\eta$ evolve in time.
According to Taylor’s frozen-turbulence hypothesis \cite{taylor1938,Tatarskii2016}, the dominant contribution to the temporal evolution arises from the transverse component of the wind velocity $v$.
In contrast, the intrinsic evolution of turbulence eddies and their longitudinal motion contribute only weakly.
As a result, the time dependence of the refractive-index fluctuations can be expressed as
	\begin{align}
		\delta n_t(\mathbf{r};z)=\delta n(x+vt,y;z),
	\end{align}
which should be used in Eq.(\ref{Eq:paraxial}) in place of $\delta n(\mathbf{r};z)$.
The corresponding solution $u_t(\mathbf{r};z)$ then yields the time-dependent transmittance,
	\begin{align}\label{Eq:eta_time}
		\eta(t) =\int_{\mathcal{A}} d^2\textbf{r} |u_t(\mathbf{r};L)|^2.
	\end{align}
A typical realization of this random process, obtained from numerical simulations, is shown in Fig.~\ref{Fig:StochasticProcess}.

Analytical techniques---for example, the phase approximation of the Huygens--Kirchhoff method \cite{Aksenov1979,Banakh1979,Mironov1977}---can be applied to evaluate important field correlation functions of the second
	\begin{align}\label{Eq:G2}
		\Gamma_2(\textbf{r};z)= \left\langle I(\textbf{r};z) \right\rangle
	\end{align}
and fourth order
	\begin{align}\label{Eq:G4}
		\Gamma_4(\textbf{r}_1, \textbf{r}_2;z)= \left\langle I(\textbf{r}_1;z)I(\textbf{r}_2;z) \right\rangle,
	\end{align}
respectively. 
These functions provide the basis for calculating several key parameters.
First, they allow us to obtain the first two moments of the transmittance,
\begin{align}\label{Eq:etaAvrg}
\left\langle\eta\right\rangle =\int_{\mathcal{A}} d^2\textbf{r} \Gamma_2 \left(\textbf{r}; L\right),
\end{align}
\begin{align}\label{Eq:eta2Avrg}
\left\langle\eta^2\right\rangle =\int_{\mathcal{A}} d^2\textbf{r}_1 d^2\textbf{r}_2 \Gamma_4 \left(\textbf{r}_1, \textbf{r}_2; L\right).
\end{align}
Second, by introducing the random beam-centroid position \cite{Fante1975}
	\begin{align}\label{Eq:centroidCoord}
		\textbf{r}_0 = \int_{\mathbb{R}^2} d^2\textbf{r} \textbf{r} I(\textbf{r};L),
	\end{align}
and assuming the coordinate system is chosen such that $\left\langle\mathbf{r}_0\right\rangle=0$, we obtain the beam-wandering variance,
	\begin{align}\label{Eq:bwVariance}
		\sigma^2_{\text{bw}} = \left\langle \Delta x_0^2 \right \rangle=\left\langle x_0^2 \right\rangle = \int_{\mathbb{R}^4} d^2\textbf{r}_1 d^2\textbf{r}_2 x_1 x_2 \Gamma_4(\textbf{r}_1, \textbf{r}_2;L).
	\end{align}
Finally, by introducing the squared instantaneous beam-spot radius \cite{Fante1975},
	\begin{align}\label{Eq:simpleS}
		S = 4\int_{\mathbb{R}^2} d^2\textbf{r} (x-x_0)^2 I(\textbf{r};L),
	\end{align}
we can derive its mean value,
	\begin{align}\label{Eq:S}
		\left\langle S \right\rangle=4\left(\int_{\mathbb{R}^2} d^2\textbf{r} x^2 \Gamma_2(\textbf{r};L) - \left\langle x_0^2 \right\rangle \right),
	\end{align}
and, after suitable approximations \cite{pechonkin2025}, its second moment,
	\begin{align}\label{Eq:S**2simplif}
		\left\langle S^2 \right\rangle \approx 16\bigg( \int_{\mathbb{R}^4} d^2\textbf{r}_1 d^2\textbf{r}_2 x_1^2 x_2^2 \Gamma_4(\textbf{r}_1, \textbf{r}_2;L)
		- \frac{1}{2}\left\langle x_0^2 \right\rangle\left\langle S \right\rangle-3\left\langle x_0^2 \right\rangle^2 \bigg).
	\end{align}
These parameters naturally enter into analytical models of the PDT.

\section{ANALYTICAL EXPRESSIONS FOR PARAMETERS}

In this section we briefly provide reader with the list of analytical expressions for the parameters \cite{pechonkin2025} $\sigma^2_{\text{bw}} $, $\left\langle S \right\rangle$, $\left\langle S^2 \right\rangle$,  $\left\langle \eta \right\rangle$, and $\left\langle \eta^2 \right\rangle$ obtained with the phase approximation of the Huygens-Kirchhoff method \cite{Aksenov1979,Banakh1979,Mironov1977} under conditions of weak impact of the turbulence.
The latter means that for the Rytov parameter the condition $\sigma_R^2=1.23C_n^2k^{7/6}L^{11/6}\ll 1$ is satisfied. 
For the beam-wandering variance and first two moments of $S$ we have
	\begin{align}\label{Eq:bwFinal}
		\sigma_{\mathrm{bw}}^2=0.31 W_0^2 \sigma_R^2\Omega^{-7/6}-0.06  W_0^2 \sigma_R^4\Omega^{-1/3},
	\end{align}
	\begin{align}\label{Eq:SFinal}
		\left\langle S \right\rangle= W_0^2\Omega^{-2}+2.93  W_0^2 \sigma_R^2\Omega^{-7/6}
		+0.24  W_0^2 	\sigma_R^4\Omega^{-1/3},
	\end{align}
	\begin{align}\label{Eq:S2Final}
		\left\langle S^2 \right\rangle= W_0^4\Omega^{-4}+6.48 W_0^4 \sigma_R^2\Omega^{-19/6}+9.40 W_0^4 \sigma_R^4\Omega^{-7/3}+2.60W_0^4 \sigma_R^6\Omega^{-3/2}-0.05W_0^4 \sigma_R^8\Omega^{-2/3},
	\end{align} 
respectively.
Here $\Omega={kW_0^2}/{2L}$ is the Fresnel number, and also we assume that the beam is focused, i.e., $L=F$.
Similarly, for the first two moments of the transmittance one can provide approximate analytical expressions
    \begin{align}\label{Eq:etaFinal}
		\left\langle\eta\right\rangle =1-\exp\left( -\frac{a^2}{0.5 W_0^2\Omega^{-2}+0.66 W_0^2\sigma_R^2\Omega^{7/6}} \right),
	\end{align}
	\begin{align}\label{Eq:eta2Final}
		\left\langle\eta^2\right\rangle =\left[ 1-\exp\left( -\frac{4a^2}{W_0^2\Omega^{-2}\left( 	1+2v\Omega^2 \right)} \right)\right]
		\left[ 1-\exp\left( -\frac{a^2 \left( 1+2v\Omega^2 \right)}{vW_0^2} \right)\right],
	\end{align}
where $v=\Omega^{-2}+3.17\sigma_R^2\Omega^{-7/6}$.

\section{PROBABILITY DISTRIBUTION OF TRANSMITTANCE}

In this section, we briefly describe the existing PDT models.
As mentioned, these models should explicitly depend only on parameters that can be analytically derived from the field correlation functions $\Gamma_2(\textbf{r};L)$ and $\Gamma_4(\textbf{r}_1, \textbf{r}_2;L)$.
We begin with empirical models and subsequently discuss those based on explicit physical assumptions regarding the beam shape and the evolution of its profile.

\subsection{Truncated Log-Normal and Beta Models} 

Historically, the log-normal distribution was the first to be used in describing quantum-light propagation through turbulent media.
Strictly speaking, however, the original considerations \cite{diament70,perina72,perina73,milonni04} referred to the intensity at a single point rather than to the transmittance.
This intensity can be defined as
\begin{equation}
I(0)=\lim_{\mathcal{A}\rightarrow 0} \frac{\eta(\mathcal{A})}{\mathcal{A}},
\end{equation}
where $\eta(\mathcal{A})$ denotes the transmittance for an aperture of area $\mathcal{A}$.
A fundamental difference between this quantity and the transmittance is that $I(0)$ is defined on the interval $[0,+\infty)$, whereas the transmittance is restricted to $\eta\in[0,1]$.
For this reason, the log-normal distribution can be consistently applied to classical electromagnetic fields.
However, it remains unclear how to extend this approach to quantum fields in a non-contradictory way.
By contrast, the transmittance admits a natural quantum description, but its domain does not coincide with that of the log-normal distribution.

\begin{figure} [ht!]
	\begin{center}
		\begin{tabular}{c} %% tabular useful for creating an array of images 
			\includegraphics[width=\linewidth]{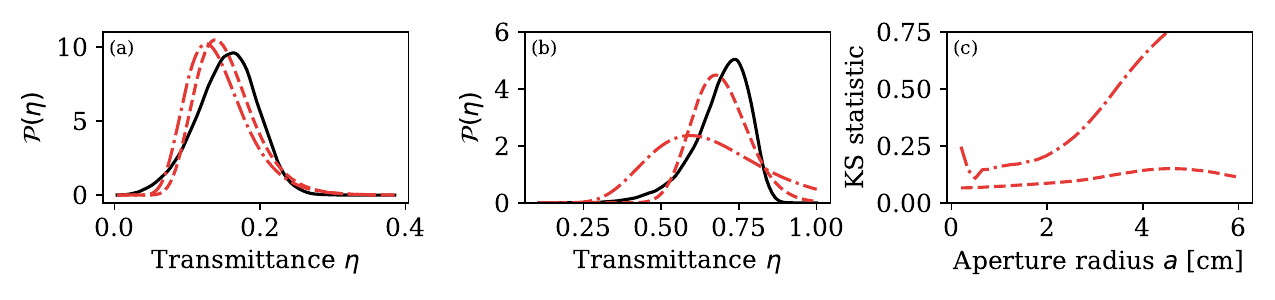}
		\end{tabular}
	\end{center}
	\caption[example] 
	%>>>> use \label inside caption to get Fig. number with \ref{}
	{ \label{Fig:LogNormal} 
		Log-normal and numerically simulated PDTs for a channel characterized by the following parameters: $C_n^2=1\times10^{-15}~\textrm{m}^{-2/3}$, $L_0=80~\textrm{m}$, $\ell_0=1~\textrm{mm}$, $W_0=2.78~\textrm{cm}$, $F_0=L$, $L=3~\textrm{km}$, and wavelength $\lambda=809~\textrm{nm}$. 
		(a) Aperture radius $a =0.9~\textrm{cm}$, where the log-normal PDT provides a satisfactory fit. (b) Aperture radius $a =2.6~\textrm{cm}$, where the log-normal PDT deviates more from the numerical simulations. (c) KS statistic between the numerical data and the log-normal PDT as a function of the aperture radius.
		Dashed and dotted-dashed lines correspond to PTDs estimated with numerically simulated and analytical parameters, respectively.}
\end{figure}

To address this issue one can truncate the log-normal distribution to the domain $\eta \in [0,1]$, provided that its tail is not extensive.
It should be emphasized, however, that the high-intensity tail can have a significant impact when postselection on the transmittance is applied, either explicitly or implicitly \cite{semenov10,gumberidze16,vasylyev16,klen2023}.
This yields the PDT in the form
		\begin{align}\label{lnTrunc}
			\mathcal{P}(\eta)= \mathcal{P}(\eta|\mu,\sigma^2)=
				\left\{\begin{array}{l c}
				\frac{1}{\mathcal{F}(1)}\frac{1}{\sqrt{2\pi}\eta\sigma}\exp\Bigl[-\frac{(\ln \eta+\mu )^2}{2\sigma^2}\Bigr],&\eta\in[0,1]\\
				0,&\text{else}
			\end{array}\right. ,
		\end{align}
where $\mathcal{F}(1)$ is the cumulative distribution function of the log-normal distribution evaluated at $\eta{=}1$.
It depends on two parameters, $\mu$ and $\sigma^2$, which can be approximately expressed in terms of the two first moments of the transmittance as
	\begin{align}\label{Mu}
		\mu=\mu(\langle\eta\rangle,\langle\eta^2\rangle)\approx-\ln\left[\frac{\langle\eta\rangle^2}{\sqrt{\langle\eta^2\rangle}}\right],
	\end{align}
	\begin{align}\label{Sigma}
		\sigma^2=\sigma^2(\langle\eta\rangle,\langle\eta^2\rangle)\approx\ln\left[\frac{\langle\eta^2\rangle}{\langle\eta\rangle^2}\right].
	\end{align}
If the tail of the distribution is not extensive, this approach provides a PDT with the clear advantage of consistently reproducing the first two moments, which are crucial for many quantum applications.
However, the overall shape of this distribution may differ significantly from those obtained numerically, see Fig.~\ref{Fig:LogNormal}.

To properly quantify the difference between numerical and analytical PDTs, we employ the Kolmogorov–Smirnov (KS) statistic \cite{Berger2014},
	\begin{align}
		D_M = \sup_\eta \left| F_M(\eta) - F(\eta) \right|.
	\end{align}
Here, $F(\eta)$ is the cumulative probability distribution corresponding to the analytical PDT, while
	\begin{align}
		F_M(\eta)=\frac{1}{M}\sum_{i=1}^M \theta(\eta-\eta_i)
	\end{align}
is the empirical cumulative distribution obtained from $M$ numerically sampled transmittances $\eta_i$ \cite{klen2023}, and $\theta(\eta)$ is the Heaviside step function.
The same measure will be applied to characterize other analytical models as well.
 
Another empirical PDT model, parameterized only by the first two moments of the transmittance, is based on the Beta distribution \cite{klen2023}.
This distribution is given by
        \begin{align}\label{Eq:pdt_beta}
			\mathcal{P}(\eta| a, b) = \frac{1}{B(a, b)} \eta^{a-1} (1-\eta)^{b-1},
		\end{align}
where $B(a, b)$ is the Beta function.
The parameters $a$ and $b$,
	\begin{align}
		&a = a\left(\langle\eta\rangle,\langle\eta^2\rangle\right)= \frac{\langle\eta\rangle - \langle\eta^2\rangle}{ \langle\eta^2\rangle- \langle\eta\rangle^2}\langle \eta \rangle,\label{Eq:a} \\
		&b =b\left(\langle\eta\rangle,\langle\eta^2\rangle\right)=  a\left(\langle\eta\rangle,\langle\eta^2\rangle\right) \left( \frac{1}{\langle\eta\rangle} - 1 \right),\label{Eq:b}
	\end{align}
are expressed in terms of $\langle\eta\rangle$ and $\langle\eta^2\rangle$.
Unlike the log-normal PDT, the Beta distribution is naturally defined on the interval $\eta \in [0,1]$, and therefore does not require artificial truncation.

\begin{figure} [ht!]
	\begin{center}
		\begin{tabular}{c} %% tabular useful for creating an array of images 
			\includegraphics[width=\linewidth]{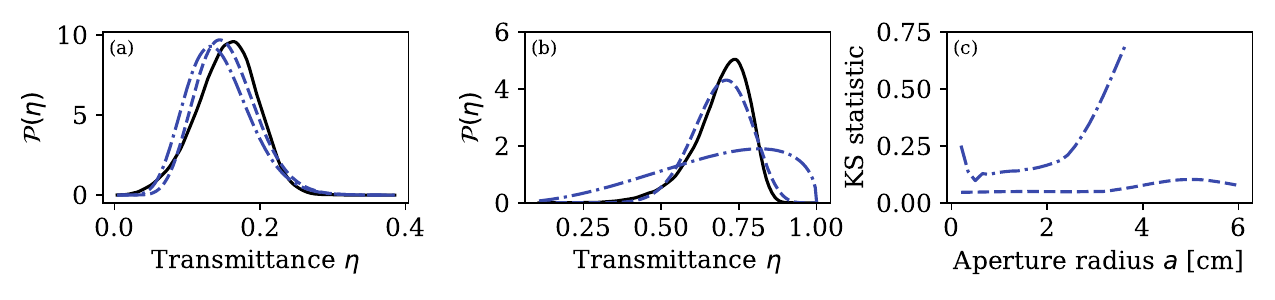}
		\end{tabular}
	\end{center}
	\caption[example] 
	%>>>> use \label inside caption to get Fig. number with \ref{}
	{ \label{Fig:Beta} 
		Beta and numerically simulated PDTs for a channel characterized by the following parameters: $C_n^2=1\times10^{-15}~\textrm{m}^{-2/3}$, $L_0=80~\textrm{m}$, $\ell_0=1~\textrm{mm}$, $W_0=2.78~\textrm{cm}$, $F_0=L$, $L=3~\textrm{km}$, and wavelength $\lambda=809~\textrm{nm}$ . (a) Aperture radius $a =0.9~\textrm{cm}$, where the Beta PDT provides a satisfactory fit. (b) Aperture radius $a =2.6~\textrm{cm}$, where the Beta PDT yields a poorer but still acceptable fit. (c) KS statistic between the numerical data and the Beta PDT as a function of the aperture radius.
		Dashed and dotted-dashed lines correspond to PTDs estimated with numerically simulated and analytical parameters, respectively.}
\end{figure} 

The advantage of the Beta PDT lies not only in being defined on the correct domain, but, more importantly, in providing a much better fit to numerical data compared to the log-normal PDT, see Fig.~\ref{Fig:Beta}.
For this reason, the Beta PDT is particularly suitable in situations where the channel must be characterized by only the first two moments of the transmittance.

In both cases, we used two methods to calculate the moments $\langle\eta\rangle$ and $\langle\eta^2\rangle$.
The first is based on the analytical expressions (\ref{Eq:etaFinal}) and (\ref{Eq:eta2Final}).
However, the corresponding results reflect not only the incompleteness of the models but also errors introduced by the analytical approximations used to derive these expressions.
To isolate the performance of the models themselves, we also calculated the moments $\langle\eta\rangle$ and $\langle\eta^2\rangle$ directly from the numerical data.
In this case, any discrepancies between the analytical and numerical PDTs are solely due to the model incompleteness.

\subsection{Beam-wandering model}

The beam-wandering model provides a simple framework in which explicit assumptions are made about the beam shape and the motion of the beam centroid \cite{vasylyev12}.
Its formulation relies on the following assumptions:
\begin{enumerate}
	\item The beam shape at the receiver remains Gaussian.
	\item The squared beam-spot radius $S$ is constant.
	\item The beam centroid follows a two-dimensional normal distribution centered at the aperture center. This assumption has been verified numerically with high accuracy \cite{klen2023}.
\end{enumerate}

Together, these assumptions lead to an explicit expression for the PDT \cite{vasylyev12}, which depends on two parameters, $\sigma_{\mathrm{bw}}^2$ and $S$,
    \begin{align}\label{Eq:BWPDT}
		\mathcal{P}(\eta|\sigma_{\mathrm{bw}}^2,S)=\frac{R^2(S)}{\sigma^2_{\text{bw}} \eta \lambda(S) } \left(\ln\frac{\eta_0(S)}{\eta} \right)^{2/\lambda(S)-1} \exp\left[ -\frac{R^2(S)}{2\sigma^2_{\text{bw}}} \left(\ln\frac{\eta_0(S)}{\eta} \right)^{2/\lambda(S)} \right]
	\end{align}
for $\eta\in\left[0, \eta_0  \right]$ and $\mathcal{P}(\eta|S)=0$ otherwise. 
Here
\begin{align}\label{Eq:eta0}
\eta_0(S)=1-\exp\left(-\frac{a^2}{S}\right)
\end{align}
is the maximal transmittance for the given $S$,	
\begin{align}\label{Eq:lambda}
\lambda(S) = 8 \frac{a^2}{S} \frac{\exp\left(-4\frac{a^2}{S} \right) \BesselM_1\left(4\frac{a^2}{S} \right)}{1-\exp\left(-4\frac{a^2}{S} \right) \BesselM_0\left(4\frac{a^2}{S} \right)}  \left[\ln\left( \frac{2\eta_0(S)}{1-\exp\left(-4\frac{a^2}{S} \right) \BesselM_0\left(4\frac{a^2}{S} \right)} \right)\right]^{-1}
\end{align}
is the shape parameter,
\begin{align}\label{Eq:R}
R(S)=a \left[\ln\left( \frac{2\eta_0(S)}{1-\exp\left(-4\frac{a^2}{S} \right) \BesselM_0\left(4\frac{a^2}{S} \right)} \right)\right]^{-1/\lambda(S)}.
\end{align}
is the scale parameter, $a$ is the aperture radius, $\BesselM_n(x)$ denotes the modified Bessel function. 
In the standard formulation of the model, one assumes $S=\langle S^2 \rangle$.

\begin{figure} [ht!]
	\begin{center}
		\begin{tabular}{c} %% tabular useful for creating an array of images 
			\includegraphics[width=\linewidth]{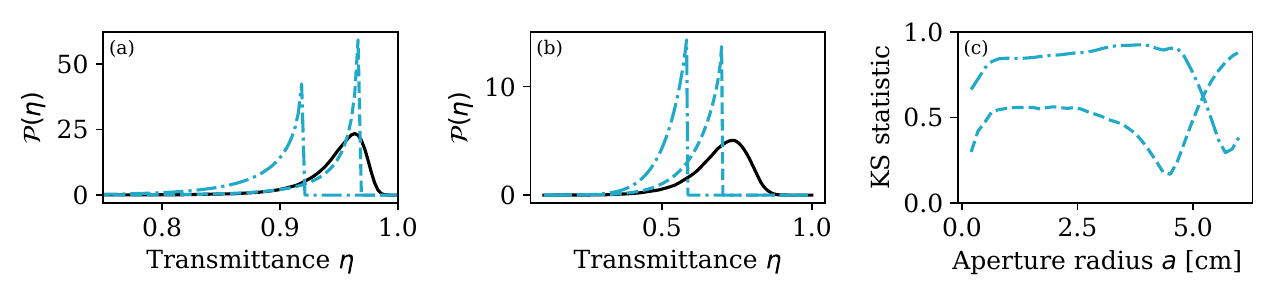}
		\end{tabular}
	\end{center}
	\caption[example] 
	%>>>> use \label inside caption to get Fig. number with \ref{}
	{ \label{Fig:BW} 
		Beam-wandering and numerically simulated PDTs for a channel characterized by the following parameters: $C_n^2=1\times10^{-15}~\textrm{m}^{-2/3}$, $L_0=80~\textrm{m}$, $\ell_0=1~\textrm{mm}$, $W_0=2.78~\textrm{cm}$, $F_0=L$, $L=3~\textrm{km}$, and wavelength $\lambda=809~\textrm{nm}$.
		(a) Aperture radius $a =4.4~\textrm{cm}$, where the beam-wandering PDT provides a satisfactory fit.
		(b) Aperture radius $a =2.6~\textrm{cm}$, where the PDT reproduces the general shape but with a shifted mode.
		(c) KS statistic between the numerical data and the beam-wandering PDT as a function of aperture radius.
		Dashed and dotted-dashed lines correspond to PTDs estimated with numerically simulated and analytical parameters, respectively.}
\end{figure}

In Fig.~\ref{Fig:BW} the beam-wandering PDT is compared with numerical simulations.
The model reproduces the overall shape of the PDT but its mode can be significantly shifted.
Agreement is achieved only when the aperture radius slightly exceeds the long-term beam radius, defined as
$W_{\mathrm{LT}}^2=\langle S^2 \rangle+4\sigma_{\mathrm{bw}}^2$.
This discrepancy originates from the assumption that the instantaneous beam profile is Gaussian.
In reality, the profile deviates strongly from Gaussianity, leading to disagreement with simulations.
Nevertheless, the beam-wandering PDT remains useful in scenarios where beam wandering arises from source jitter rather than atmospheric turbulence. 

A natural way to improve the model is to match the parameters $S$ and $\sigma_{\mathrm{bw}}^2$ to the first two transmittance moments, $\langle\eta\rangle$ and $\langle\eta^2\rangle$.
Using the result of Esposito \cite{Esposito1967}, the moments of the distribution (\ref{Eq:BWPDT}) can be written as
	\begin{align}\label{Eq:EtaBW}
		\left<\eta\right>_{S,\sigma_{\mathrm{bw}}^2} = 1 - \exp\!\left(-2\frac{a^2}{4\sigma_{\mathrm{bw}}^2 + S}\right),
	\end{align} 
	\begin{align}\label{Eq:Eta2BW}
		\left<\eta^2\right>_{S,\sigma_{\mathrm{bw}}^2} &= 1-2 \exp\! \left(-2\frac{a^2}{4\sigma_{\mathrm{bw}}^2 + S}\right) \\&+  
		\exp\! \left(-\frac{\alpha^2}{2}\right)\left[1-Q\!\left(\frac{\alpha}{\sqrt{1-\beta^2}}, \frac{\alpha \beta}{\sqrt{1-\beta^2}}\right) \right.\nonumber
	+ \left.Q\!\left(\frac{\alpha \beta}{\sqrt{1-\beta^2}}, \frac{\alpha}{\sqrt{1-\beta^2}}\right)\right],\nonumber
	\end{align}
where 
	\begin{align}
		\alpha=\frac{2a}{\sqrt{S}}\left[\frac{2 p(p+1)}{2 p^2+3 p+1}\right]^{1 / 2}, 
	\end{align} 
	\begin{align}
		\beta = (2p+1)^{-1}, \quad p = \frac{1}{8}\frac{S}{\sigma_{\mathrm{bw}}^2},
	\end{align} 
and $Q(x,y)$ is the Marcum $Q$-function of the first order \cite{Marcum_book,Agrest_book,Vasylyev2013}.
By substituting $\langle\eta\rangle$ and $\langle\eta^2\rangle$ from Eqs.~~(\ref{Eq:etaAvrg}) and (\ref{Eq:eta2Avrg}) into the left-hand sides of Eqs.~(\ref{Eq:EtaBW}) and (\ref{Eq:Eta2BW}), one obtains a system of algebraic equations for $S$ and $\sigma_{\mathrm{bw}}^2$.
Numerical solution of this system yields parameter values that differ from those defined by Eqs.~(\ref{Eq:bwVariance}) and (\ref{Eq:S}), but restore agreement with the mode of the numerical PDT.
Although this method provides a reasonable empirical correction, it cannot be regarded as a consistent procedure for estimating $\langle S \rangle$ and $\sigma_{\mathrm{bw}}^2$, since the non-Gaussian beam profile introduces significant bias in many cases.

\subsection{Circular-beam model}

The circular-beam model \cite{pechonkin2025} provides a natural extension of the beam-wandering model to a more realistic scenario.
It is based on the following assumptions:
\begin{enumerate}
	\item The beam shape at the receiver remains circular Gaussian.
	\item The squared beam-spot radius $S$ fluctuates according to the log-normal law,
		\begin{align}\label{Eq:lognorm}
			P(S|\mu_S,\sigma_S^2)=\frac{1}{S\sigma_S\sqrt{2\pi}}\exp{-\frac{\left(\ln{S}-\mu_S\right)^2}{2\sigma_S^2}},
		\end{align}
	which is supported by numerical simulations \cite{pechonkin2025}.
	Here $\mu_S$ and $\sigma_S^2$ are model parameters.
	\item The beam centroid follows a two-dimensional normal distribution centered at the aperture. This assumption has been numerically verified with high accuracy \cite{klen2023}.
	\item The squared beam-spot radius $S$ and the beam-centroid position $\mathbf{r}_0$ are statistically independent.
\end{enumerate}

With these assumptions, the PDT takes the form
	\begin{align}\label{Eq:fullprob}
		\mathcal{P}(\eta|\mu_S,\sigma_S^2)=\int_{0}^\infty dS \mathcal{P}(\eta|S) P(S|\mu_S,\sigma_S^2).
	\end{align}
The parameters $\mu_S$ and $\sigma_S^2$ are related to the moments of $S$ as
	\begin{align}\label{Eq:mu}
		\mu_S=\ln{\left(\frac{\left\langle S \right\rangle^2}{\sqrt{\left\langle S^2 \right\rangle}}\right)},
	\end{align}
	\begin{align}\label{Eq:sigma}
		\sigma_S^2=\ln{\left(\frac{\left\langle S^2 \right\rangle}{\left\langle S \right\rangle^2}\right)}.
	\end{align}
These moments are, in turn, connected to the field-correlation functions $\Gamma_2(\textbf{r};L)$ and $\Gamma_4(\textbf{r}_1, \textbf{r}_2;L)$ via Eqs.~(\ref{Eq:S}) and (\ref{Eq:S**2simplif}).

\begin{figure} [ht!]
	\begin{center}
		\begin{tabular}{c} %% tabular useful for creating an array of images 
			\includegraphics[width=\linewidth]{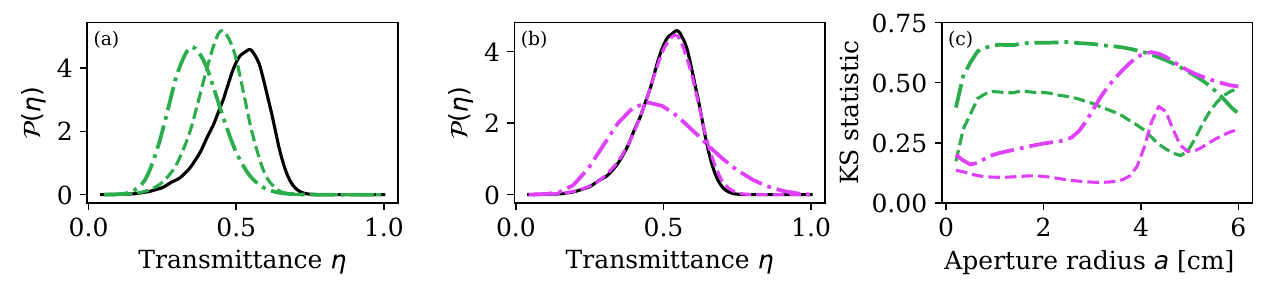}
		\end{tabular}
	\end{center}
	\caption[example] 
	%>>>> use \label inside caption to get Fig. number with \ref{}
	{ \label{Fig:CBM} 
		Circular-beam and numerically simulated PDTs for a channel characterized by the following parameters: $C_n^2=1\times10^{-15}~\textrm{m}^{-2/3}$, $L_0=80~\textrm{m}$, $\ell_0=1~\textrm{mm}$, $W_0=2.78~\textrm{cm}$, $F_0=L$, $L=3~\textrm{km}$, wavelength $\lambda=809~\textrm{nm}$, and $a =2~\textrm{cm}$.
		(a) Parameters matched to the moments of $S$.
		(b) Parameters matched to the moments of $\eta$.
		(c) KS statistic between the numerical data and the circular-beam PDT as a function of the aperture radius.
		Green and violate colors correspond to matching by moments of $S$ and by moments of $\eta$, respectively. 
		Dashed and dotted-dashed lines correspond to PTDs estimated with numerically simulated and analytical parameters, respectively.}
\end{figure} 

In this formulation, the circular-beam PDT suffers from problems similar to those of the beam-wandering PDT: even if the overall shape resembles the numerical data, the mode of the distribution can be significantly biased.
This discrepancy arises from the strong non-Gaussianity of the instantaneous beam profile.
The model can be improved in a way similar to the beam-wandering PDT—by matching the parameters $\mu_S$ and $\sigma_S^2$ to the transmittance moments $\langle\eta\rangle$ and $\langle\eta^2\rangle$.
In this case, the beam-wandering variance $\sigma_{\mathrm{bw}}^2$ remains the same as obtained from Eq.~(\ref{Eq:bwVariance}).
From Eq.~(\ref{Eq:fullprob}), the first two moments of the transmittance can be expressed in terms of the conditional moments $\left<\eta\right>_{S,\sigma_{\mathrm{bw}}^2}$ and $\left<\eta^2\right>_{S,\sigma_{\mathrm{bw}}^2}$, cf. Eqs.~(\ref{Eq:EtaBW}) and (\ref{Eq:Eta2BW}),
	\begin{align}
		&\left\langle \eta \right\rangle=\int_{0}^{+\infty}dS  P(S|\mu_S,\sigma_S^2)\left\langle\eta\right\rangle_{S,\sigma_{\mathrm{bw}}^2},\label{Eq:EqEta1}\\
		&\left\langle \eta^2 \right\rangle=\int_{0}^{+\infty}dS P(S|\mu_S,\sigma_S^2)\left\langle\eta^2\right\rangle_{S,\sigma_{\mathrm{bw}}^2}.\label{Eq:EqEta2}
	\end{align}
These equations form a system of algebraic relations.
By solving them numerically for $\mu_S$ and $\sigma_S^2$, one obtains parameter values that correct the problem of mode displacement.
As shown in Fig.~\ref{Fig:CBM}, this technique yields a satisfactory fit to the numerical data.

\subsection{Elliptic-beam model}

The elliptic-beam model \cite{vasylyev16,vasylyev17} represents the next level of complexity toward a more realistic description.
It is based on assumptions similar to those of the circular-beam approximation.
However, it requires more involved analytical and numerical calculations.

The key difference from the circular-beam model is that, in the elliptic-beam model, the beam profile at the receiver is Gaussian but elliptically shaped.
Both the ellipse axes and their orientation are random variables. 
Technically, the corresponding PDT is given by
	\begin{align}\label{Eq:EllBPDT}
		\mathcal{P}\left(\eta|\sigma_{\mathrm{bw}}^2,\mu_S,\boldsymbol{\Sigma}\right)=\int_{\mathbb{R}^2}d^2\mathbf{r}\,\rho_{\mathbf{G}}(\mathbf{r}_0|\sigma_{\mathrm{bw}}^2)
		\int_{\Omega} d \mathbf{S} \,\rho_S(\mathbf{S}|\mu_S,\boldsymbol{\Sigma})
		\delta\left[\eta-\eta(\mathbf{r}_0,\mathbf{S})\right].
	\end{align}
Here $\rho_{\mathbf{G}}(\mathbf{r}_0|\sigma_{\mathrm{bw}}^2)$ is the normal distribution of the beam-centroid coordinates with variance $\sigma_{\mathrm{bw}}^2$,
	 \begin{align}\label{Eq:MatrixS_Definition}
		 \mathbf{S}=\left( \begin{array}{cc}
		 S_{xx} & S_{xy}  \\
		 S_{xy} & S_{yy} \end{array} \right)
		 =4\int_{\mathbb{R}^2}d^2\mathbf{r}
		 \,\left[(\mathbf{r}-\mathbf{r}_0)(\mathbf{r}-\mathbf{r}_0)^\mathrm{T}
		 \right ] \,
		 I(\mathbf{r},z)
	 \end{align}		
is the instantaneous spot-shape matrix.
This matrix is positive-semidefinite and symmetric, distributed according to $\rho_S(\mathbf{S}|\mu_{S},\boldsymbol{\Sigma})$.
 The function $\eta(\mathbf{r}_0,S)$ denotes the transmittance efficiency trough the aperture for a beam with centroid coordinates $\mathbf{r}_0$ and the spot-shape matrix $\mathbf{S}$.
The distribution $\rho_S(\mathbf{S}|\mu_{S},\boldsymbol{\Sigma})$ depends on the parameters $\mu_{S}$ and $\boldsymbol{\Sigma}$, specified below. 

Let $W_1^2$ and $W_2^2$ be the eigenvalues of the spot-shape matrix $\mathbf{S}$, i.e., squared ellipse semi-axes.
The elements of $\mathbf{S}$ can then be expressed in terms of $W_1^2$, $W_2^2$, and the angle of ellipse orientation, $\phi$, as
\begin{align}
&S_{xx}=W_1^2\cos^2\!\phi+W_2^2\sin^2\!\phi,\label{Eq:Sxx}\\
&S_{yy}=W_1^2\sin^2\!\phi+W_2^2\cos^2\!\phi,\label{Eq:Syy}\\
&S_{xy}=S_{yx}=\frac{1}{2}\Bigl(W_1^2-W_2^2\Bigr)\sin2\phi.\label{WxWyrelation}
\end{align} 
Instead of $W_i^2$, we introduce the logarithmic variables
	\begin{align}
		\Theta_i=\ln W_i^2, \quad i=1,2.
	\end{align}
They are assumed to follow a two-fold symmetric Gaussian distribution with mean vector $\begin{pmatrix}
\mu_{S} & \mu_{S} \end{pmatrix}^{\mathrm{T}}$ and covariance matrix $\boldsymbol{\Sigma}_{i,j}=\langle \Delta\Theta_i\Delta\Theta_j\rangle$.
Here $\mu_S$ is the same parameter as in the circular-beam model, defined by Eq.~(\ref{Eq:mu}).
The elements of $\boldsymbol{\Sigma}$ are related to the moments of $W_i^2$ as
	\begin{align}\label{Eq:ThetaCovariances}
		\langle \Delta\Theta_i\Delta\Theta_j\rangle=
		\ln\left[
		\frac{\langle W_i^2 W_j^2\rangle}{\langle	
			W_i^2\rangle\langle	
			W_j^2\rangle}\right].
	\end{align}
The averaged squared ellipse semi-axes satisfy $\langle W_1^2\rangle=\langle W_2^2\rangle=\langle S\rangle$ and are related to the field-correlation functions $\Gamma_2(\textbf{r};L)$ and $\Gamma_4(\textbf{r}_1, \textbf{r}_2;L)$  through Eq.~(\ref{Eq:S}).
The second moments $\langle W_{i}^2W_{j}^2\rangle$ are expressed in terms of these functions as
	\begin{align}
		\langle
		W_{i}^2W_{j}^2\rangle=8\left[{-}8\,\delta_{ij}\langle
		x_0^2\rangle^2
		{-}\langle x_0^2\rangle \langle
		W_{i}^2\rangle
		\label{Eq:W12ViaGamma}
		+\int_{\mathbb{R}^4}d^4\mathbf{r}
		\,
		\left[x_1^2x_2^2\left(4\delta_{ij}{-}1\right)-x_1^2y_2^2\left(4\delta_{ij}{-}3\right)\right]
		\Gamma_4\!\left(\mathbf{r}_1,\mathbf{r}_2;L\right)
		\right].
	\end{align}
For the orientation angle $\phi$, we assume a uniform distribution over $(0,\pi/2]$.
Thus, the distribution $\rho_S(\mathbf{S}|\mu_S,\boldsymbol{\Sigma})$ is parameterized by three quantities: $\mu_S$, $\langle \Delta\Theta_1^2\rangle$, and $\langle \Delta\Theta_1\Delta\Theta_2\rangle$.
It is Gaussian in $\Theta_1$ and $\Theta_2$ and uniform in $\phi$.
The set $\Omega$ is defined as $\{\Theta_1\in \mathbb{R},\Theta_2\in \mathbb{R},\phi\in(0,\pi/2]\}$.

An approximate expression for the transmittance $\eta(\mathbf{r}_0,\mathbf{S})$ in Eq.~(\ref{Eq:EllBPDT}) is given by
	\begin{align}
		\eta=\eta_{0} \exp\left\{-\left[\frac{r_0/a}
		{R\left(W_{\rm
					eff}^2\left(\phi{-}\varphi_0\right)\right)}\right]^{\lambda\bigl(W_{\rm
					eff}^2\left(\phi{-}\varphi_0\right)\bigr)}\right\}.\label{Tapprox}
	\end{align}
Here
	\begin{align}
		W_\textrm{eff}^2\left(\chi\right)&{=}4a^2\left[\mathcal{W}\Bigl(\frac{4a^2}{
			W_1W_2} e^
		{\frac{a^2}{W_1^2}\bigl\{1+2\cos^2\!\chi\bigr\}}
		e^
		{\frac{a^2}{W_2^2}\bigl\{1+2\sin^2\!\chi\bigr\}}\Bigl)\right]^{-1},\label{Weff}
	\end{align}
where $\mathcal{W}(\xi)$ is the Lambert $W$ function,
	\begin{align}
		\eta_{0}{=}1{-}\BesselM_0\Bigl(a^2\Bigl[\frac{1}{W_1^2}{-}\frac{1}{W_2^2}\Bigr]\Bigr)e^{-a^2\bigl[\frac{1}{W_1^2}{+}\frac{1}{W_2^2}\bigr]}{-}2\left[1{-}e^{-\frac{a^2}{2}\!\bigl(\frac{1}{W_1}{-}\frac{1}{W_2}\bigr)^{2}}\!\right]
		\exp\!\left\{\!{-}\Biggl[\!\frac{\frac{(W_1+W_2)^2}{|W_1^2-W_2^2|}}{R\left(\frac{4W_1^2W_2^2}{(W_1-W_2)^2}\right)}\!\Biggr]
		^{\!\lambda\left(\frac{4W_1^2W_2^2}{(W_1-W_2)^2}\right)}\right\},
	\end{align}
$R(\xi)$ and $\lambda(\xi)$ are  scale and shape parameters defined by Eqs.~(\ref{Eq:R}) and (\ref{Eq:lambda}), respectively.
The variables $r_0$ and $\varphi_0$ denote the magnitude and orientation of the centroid vector $\mathbf{r}_0$, respectively.
 
In practice, the elliptic-beam PDT is obtained using Monte Carlo integration of Eq.~(\ref{Eq:EllBPDT}).
Random values of $\mathbf{r}_0$, $\Theta_1$, $\Theta_2$, and $\phi$ are generated, substituted into Eq.(\ref{Tapprox}).
 This procedure is repeated many times to form a sampling set of transmittance values.
This set can then be used to estimate the PDT, for example, via a histogram or a kernel-smoothing method.
As with the beam-wandering and circular-beam models, the elliptic-beam model often predicts a biased distribution mode \cite{klen2023}.
However, improving this model by matching transmittance moments (as is possible for the other two models) is generally impractical due to the heavy analytical and numerical workload involved.

\subsection{Models based on the law of total probability}

Models based on the law of total probability \cite{vasylyev18} can be regarded as modifications of the circular- and elliptic-beam approaches.
As in those approximations, they explicitly account for the beam-wandering effect.
However, unlike the previous models, they do not rely on physical assumptions about the beam shape.
Instead, the effects of beam-spot distortion are empirically described using log-normal or Beta distributions.
A key advantage of these models is that they automatically match the first two transmittance moments, which removes the need for additional improvement procedures.
On the other hand, they require explicit assumptions to ensure that analytical and numerical calculations remain tractable.

Let us consider the instantaneous intensity at the aperture plane, $I(\mathbf{r},L)$.
The beam centroid is located at $\mathbf{r}_0$, cf.Eq.(\ref{Eq:centroidCoord}).
We then introduce the intensity $I^{(c)}(\mathbf{r}^\prime,L)$ in the instantaneous coordinate frame through the relation
	\begin{align}
		I(\mathbf{r},L)=I^{(c)}(\mathbf{r}-\mathbf{r}_0,L).
	\end{align}
Our first assumption is that the random function $I^{(c)}(\mathbf{r}^\prime,L)$ is statistically independent of the beam-centroid position $\mathbf{r}_0$.
This leads to
   		\begin{align} \label{ConditGamma2}
   			\Gamma_2(\mathbf{r},L)=\int_{\mathbb{R}^2} d^2\boldsymbol{r}_0  		\rho_{\mathbf{G}}(\mathbf{r}_0|\sigma_{\mathrm{bw}}^2)\Gamma_{2}^{(c)}(\mathbf{r}-\mathbf{r}_0,L),
  		 \end{align}
where $\Gamma_2^{(c)}(\mathbf{r},L)=\langle I^{(c)}(\mathbf{r},L)\rangle$ is the second-order correlation function of the perfectly-tracked beam.
In a similar manner, the fourth-order correlation function of the perfectly tracked beam can be written as
		\begin{align}\label{ConditGamma4}
			\Gamma_4(\mathbf{R},\boldsymbol{\rho},L){=}\int_{\mathbb{R}^2} d^2\mathbf{r}_0  \rho_{\mathbf{G}}(\mathbf{r}_0|\sigma_{\mathrm{bw}}^2)\,
			\Gamma_{4}^{(c)}(\mathbf{R}{-}\sqrt{2}\mathbf{r}_0,\boldsymbol{\rho},L),
		\end{align}
where we use the coordinates $ \boldsymbol{R}{=}(\boldsymbol{r}_1{+}\boldsymbol{r}_2)/\sqrt{2}$, $\boldsymbol{\rho}{=}(\mathbf{r}_1{-}\mathbf{r}_2)/\sqrt{2}$. 
Assuming that $\rho_{\mathbf{G}}(\mathbf{r}_0|\sigma_{\mathrm{bw}}^2)$ is a Gaussian distribution with the variance $\sigma_{\mathrm{bw}}^2$, these expressions can be inverted as
	\begin{align}\label{DiffGamma2}
		\Gamma_{2}^{(c)}(\mathbf{r},L)=\exp\left[-\frac{\sigma_{\mathrm{bw}}^2}{2}\Delta_{\mathbf{r}}\right]\Gamma_2(\mathbf{r},L),
	\end{align}
	\begin{align}\label{DiffGamma4}
		\Gamma_{4}^{(c)}(\mathbf{R},\boldsymbol{\rho},L)=\exp\left[-\sigma_{\mathrm{bw}}^2\Delta_{\mathbf{R}}\right]
		\Gamma_{4}(\mathbf{R},\boldsymbol{\rho},L),
	\end{align}
where $\Delta_{\mathbf{r}}{=}\frac{\partial^2}{\partial x^2}{+}\frac{\partial^2}{\partial y^2}$ is the Laplace operator.
Therefore, Eqs.~(\ref{DiffGamma2}) and (\ref{DiffGamma4}) provide a way to obtain the field-correlation functions $\Gamma_{2}^{(c)}(\mathbf{r},L)$ and $\Gamma_{4}^{(c)}(\mathbf{R},\boldsymbol{\rho},L)$ with the beam-wandering effect removed.

If the field-correlation functions $\Gamma_{2}^{(c)}(\mathbf{r},L)$ and $\Gamma_{4}^{(c)}(\mathbf{R},\boldsymbol{\rho},L)$  are known, the first two moments of the transmittance for the perfectly tracked beam can be determined as
	\begin{align}\label{MEtar}
		\langle\eta\rangle_{r_0}=\int_{\mathcal{A}(\mathbf{r}_0)}d^2\mathbf{r} \Gamma_2^{(c)}(\mathbf{r},L),
	\end{align}
	\begin{align}\label{MEta2r}
		\langle\eta^2\rangle_{r_0}=\int_{\mathcal{A}(\mathbf{r}_0)}d^2\mathbf{r}_1
		\int_{\mathcal{A}(\mathbf{r}_0)}d^2\mathbf{r}_2 \Gamma_4^{(c)}(\mathbf{r}_1,\mathbf{r}_2,L),
	\end{align}
where $\mathcal{A}({\mathbf{r}}_0)$ denotes the aperture opening with its center displaced by $\mathbf{r}_0$.
Our next assumption is that the PDT, conditioned on the beam centroid being located at a distance $r_0$ from the aperture center, is described either by the log-normal distribution \cite{vasylyev18} or by the Beta distribution \cite{klen2023}.
In the first case, we denote it $\mathcal{P}(\eta|\mu_{r_0},\sigma_{r_0})$, where the parameters are given by $\mu_{r_0}=\mu(\langle\eta\rangle_{r_0},\langle\eta^2\rangle_{r_0})$, cf. Eq.~(\ref{Mu}) and $\sigma_{r_0}=\sigma(\langle\eta\rangle_{r_0},\langle\eta^2\rangle_{r_0})$, cf. Eq.~(\ref{Sigma}).
In the second case, we denote it as $\mathcal{P}(\eta|a_{r_0},b_{r_0})$ with the parameters $a_{r_0}=a(\langle\eta\rangle_{r_0},\langle\eta^2\rangle_{r_0})$, cf. Eq.~(\ref{Eq:a}) and $b_{r_0}=b(\langle\eta\rangle_{r_0},\langle\eta^2\rangle_{r_0})$, cf. Eq.~(\ref{Eq:b}) .
Finally, the PDT based on the law of total probability is defined as
	\begin{align}
		\mathcal{P}(\eta)=\int_{\mathbb{R}^2}  d^2 \mathbf{r}_0 \mathcal{P}(\eta|\mu_{r_0},\sigma_{r_0})\rho_{\mathbf{G}}(\mathbf{r}_0|\sigma_{\mathrm{bw}}^2)\label{Eq:Bayesian_LN}
	\end{align}	
for the log-normal case and 
	\begin{align}
		\mathcal{P}(\eta)=\int_{\mathbb{R}^2}  d^2 \mathbf{r}_0 \mathcal{P}(\eta|a_{r_0},b_{r_0})\rho_{\mathbf{G}}(\mathbf{r}_0|\sigma_{\mathrm{bw}}^2)\label{Eq:Bayesian_Beta}
	\end{align}	
for the Beta case.

The drawback of the described method is that it requires involved numerical calculations to determine the field-correlation functions $\Gamma_{2}^{(c)}(\mathbf{r},L)$ and $\Gamma_{4}^{(c)}(\mathbf{R},\boldsymbol{\rho},L)$, followed by their integration.
As an alternative, one may introduce simplifying assumptions that considerably reduce the computational effort, but at the cost of introducing additional errors in the final PDT.
A key requirement for such approximations is that the resulting PDT remains consistent with the first two moments of the transmittance.

The main assumption behind this approximation is that the conditional moments $\langle\eta\rangle_{r_0}$ and $\langle\eta^2\rangle_{r_0}$ are assumed to depend on $r_0$ in a way similar to the dependence of $\eta$ on $r_0$ for the Gaussian beam \cite{vasylyev12}.
This leads to
	\begin{align}\label{CondEta1}
		\langle\eta\rangle_{r_0}=\eta_0\exp\left[-\left(\frac{r_0}{R(\langle S\rangle)}\right)^{\lambda(\langle S\rangle)}\right],
	\end{align}
	\begin{align}\label{CondEta2}
		\langle\eta^2\rangle_{r_0}=\zeta_{0}^2\exp\left[-2\left(\frac{r_0}{R(\langle S\rangle)}\right)^{\lambda(\langle S\rangle)}\right].
	\end{align}
Here $R(\xi)$ and $\lambda(\xi)$ are  scale and shape parameters defined by Eqs.~(\ref{Eq:R}) and (\ref{Eq:lambda}), respectively.
The parameters $\eta_0$ and $\zeta_{0}^2$ are chosen such that the transmittance moments of the PDT match the values $\langle \eta\rangle$ and $\langle \eta^2\rangle$ defined by Eqs.(\ref{Eq:etaAvrg}) and (\ref{Eq:eta2Avrg}).
This requirement is satisfied if these parameters are obtained from
	\begin{align}\label{Eta1Defin}
		\eta_0=\frac{\langle\eta\rangle}{\displaystyle{\int_{0}^\infty  d\xi\,\xi \,e^{-\frac{\xi^2}{2}}e^{-\left(\frac{\sigma_{\mathrm{bw}}}{R(\langle S\rangle)}\xi\right)^{\lambda(\langle S\rangle)}}}}
	\end{align}  
and
	\begin{align}\label{Eta2Defin}
		\zeta_0^2=\frac{\langle\eta^2\rangle}{\displaystyle{\int_{0}^\infty  d\xi\,\xi \,e^{-\frac{\xi^2}{2}}e^{-2\left(\frac{\sigma_{\mathrm{bw}}}{R(\langle S\rangle)}\xi\right)^{\lambda(\langle S\rangle)}}}}.
	\end{align}
The conditional moments $\langle\eta\rangle_{r_0}$ and $\langle\eta^2\rangle_{r_0}$ determined from Eqs.~(\ref{CondEta1}) and (\ref{CondEta2}) are then used to calculate the parameters of the conditional PDT, $\mu_{r_0}$ and $\sigma_{r_0}$ or $a_{r_0}$ and $b_{r_0}$.
Finally, this conditional PDT is substituted into Eq.~(\ref{Eq:Bayesian_Beta}) to obtain the total PDT.

\section{TIME CORRELATIONS}

As mentioned earlier, realistic experiments always involve time averaging rather than the ensemble averaging considered in the previous section.
Ergodicity---i.e., the equivalence of these two averaging methods---is possible only for time intervals that significantly exceed the correlation time.
Time correlations are also crucial for the implementation of various communication protocols \cite{Klen2024}, such as enlarging the effective Hilbert-space dimension via time-bin encoding, adaptive real-time channel selection using classical light pulses, and so on.

To fully characterize the random process $\eta(t)$, one needs to specify its probability density functional $\mathcal{P}[\eta(t)]$, or, when considering discrete times, the joint probability distribution $\mathcal{P}(\eta_1,\eta_2,\ldots)$.
Here, $\eta_i=\eta(t_i)$ with $t_1<t_2<\ldots$.
In this work, we focus on two-time correlations, i.e., the joint PDT $\mathcal{P}(\eta_1,\eta_2)$.
The single-time PDT considered in the previous section is related to the two-time PDT by
 \begin{align}
 	\mathcal{P}(\eta_1)=\int_0^1d\eta_2 \mathcal{P}(\eta_1,\eta_2).
 \end{align}
The two-time PDT also provides information about the correlation function
	\begin{align}\label{Eq:CorrFunc}
		G(\tau)=\frac{\left\langle\Delta\eta(t)\Delta\eta(t+\tau)\right\rangle}{\sqrt{\left\langle\Delta\eta^2(t)\right\rangle\left\langle\Delta\eta^2(t+\tau)\right\rangle}}.
	\end{align}
As an example, let us consider adaptive real-time channel selection. 
Here, the channel transmittance is probed at time $t$ by a classical pulse, and only the events with $\eta(t)\geq\eta_{\mathrm{min}}$ are selected for sending a nonclassical pulse at a later time $t+\tau$.
This scenario is described by the conditional PDT
	\begin{align}
		\mathcal{P}\Big(\eta(t+\tau)\Big|\eta(t)\geq\eta_{\mathrm{min}}\Big)=\frac{\displaystyle\int_{\eta_{\mathrm{min}}}^1d\eta_1\mathcal{P}\Big(\eta_1,\eta(t+\tau)\Big)\Big|_{\eta_1=\eta(t)}}{\displaystyle\int_{\eta_{\mathrm{min}}}^1d\eta_1\mathcal{P}\Big(\eta_1\Big)},
	\end{align}
which is directly related to the two-time PDT.

Apart from trivial empirical scenarios, analytical models for the two-time PDT are not easily formulated, even for extensions of simple cases such as the beam-wandering model.
For this reason, a straightforward approach is to rely on numerical simulations.
As already noted in Sec.~\ref{Sec:ClassRad}, the time evolution of the transmittance is mainly driven by the transverse component of the wind velocity.
Accordingly, numerical simulations \cite{Klen2024} based on the sparse-spectrum model \cite{Charnotskii2013a,Charnotskii2013b,Charnotskii2020} of the phase-screen method \cite{Fleck1976,Frehlich2000,Lukin_book,Schmidt_book} are performed as follows:
the field amplitude is simulated at the aperture plane, the transmitted intensity fraction is calculated, the phase screens are shifted by a distance corresponding to the time $\tau$, and the procedure is repeated.
By repeating this process many times, one obtains a sample set of random pairs $(\eta_1,\eta_2)$, which can then be used to estimate the two-time PDT.

\begin{figure} [ht!]
	\begin{center}
		\begin{tabular}{c} %% tabular useful for creating an array of images 
			\includegraphics[width=\linewidth]{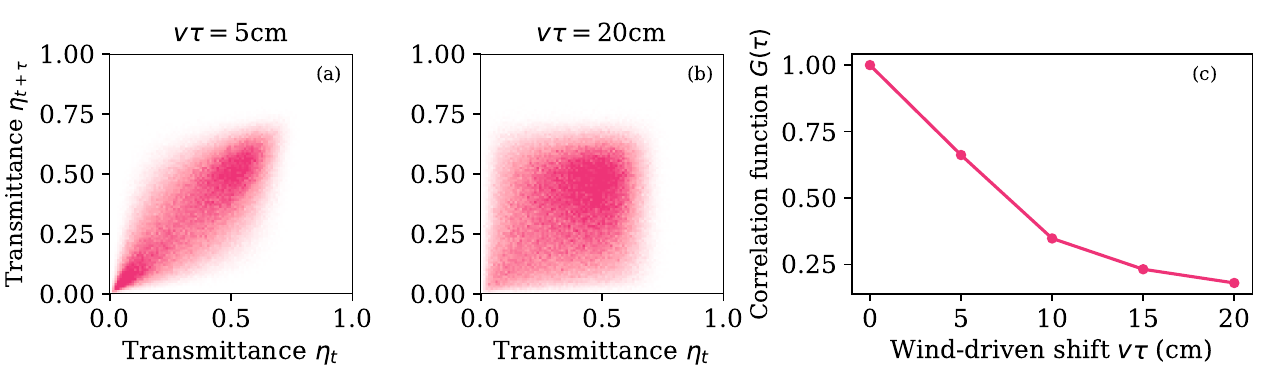}
		\end{tabular}
	\end{center}
	\caption[example] 
	%>>>> use \label inside caption to get Fig. number with \ref{}
	{ \label{Fig:2TimePDT} 
		Scatter plots of the two-time PDTs obtained from numerical simulations for a channel with the following parameters: $C_n^2=5\times10^{-15}~\textrm{m}^{-2/3}$, $L_0=1~\textrm{km}$, $\ell_0=1~\textrm{mm}$, $W_0=5.08~\textrm{cm}$, $F_0=+\infty$, $L=10.2~\textrm{km}$, and wavelength $\lambda=808~\textrm{nm}$.
		(a) $\tau = 5~\textrm{ms}$ (wind velocity $v=10~m/s$).
		(b) $\tau = 20~\textrm{ms}$ (wind velocity $v=10~m/s$).
		(c) Correlation function $G(\tau)$, cf. Eq.~(\ref{Eq:CorrFunc}).}
\end{figure}

In Fig.~\ref{Fig:2TimePDT}, we present the results of numerical simulations of the two-time PDT for two different values of $\tau$, along with the correlation function $G(\tau)$ given by Eq.~(\ref{Eq:CorrFunc}).
As expected, correlations are strong for small $\tau$ and then decrease as $\tau$ increases.
Remarkably, the correlation time is sufficiently long to enable the implementation of various communication protocols.
Moreover, these classical correlations can be directly connected to quantum correlations when, for example, two entangled pulses are transmitted through the atmospheric channel \cite{Klen2024}.

\section{SUMMARY AND DISCUSSIONS}

To summarize, the theoretical description of free-space quantum channels is a non-trivial task that must account not only for the quantum electromagnetic radiation itself but also for the methods of quantum-information encoding, as well as the detection and postprocessing techniques employed.
In the most general case, the parameters describing the free-space channel evolve randomly in time, forming a non-trivial stochastic process with finite correlation and ergodicity times.
Therefore, ensemble averaging of the quantum states transmitted through the atmosphere should replace time averaging only in physical scenarios where such a substitution is justified.
 
In this paper, we focus on theoretical models describing a widely used scenario in which quantum information is encoded in the quantum state of a quasi-monochromatic mode prepared as a pulsed Gaussian beam.
In this case, the spatial structure of the light mode at the receiver station is irrelevant, since the detection system is sensitive only to the total fraction of light intensity transmitted through the receiver aperture.
Similarly, if the detection system is not sensitive to the temporal structure of the light mode, that aspect can also be neglected.
However, photocounting techniques that register the number of photocurrent pulses within a measurement time window---and are affected by detector dead time and/or relaxation time---typically require an explicit treatment of the temporal mode structure.

In the considered scenario, the stochastic process is described solely by the random time evolution of a single quantity—the transmittance through the receiver aperture, $\eta(t)$.
A complete characterization of this process requires the probability density functional $\mathcal{P}\!\left[\eta(t)\right]$.
A less comprehensive, yet still informative, approach is based on the two-time probability density $\mathcal{P}(\eta_1,\eta_2)$, where $\eta_i=\eta(t_i)$ with $t_1<t_2$.
This distribution encodes information about the time-correlation function, which is particularly important for determining the time interval over which transmittance values can be regarded as statistically independent.
 
In most cases, analytical models for $\mathcal{P}(\eta_1,\eta_2)$ involve complicated mathematical constructions.
For this reason, we have relied on numerical simulations to address the problem.
Our results show that the correlation time in atmospheric channels is quite large compared with typical time scales in quantum optics.
On the one hand, this opens promising perspectives for implementing quantum-transfer protocols, including time-bin encoding of multimode quantum states and real-time adaptive protocols.
On the other hand, it implies that relatively long observation times are required when replacing time averaging with ensemble averaging.

For scenarios where ensemble averaging is applicable, the effect of the atmosphere on quantum states of the light mode can be described using the single-time PDT.
The most accurate method in this case is again provided by numerical simulations with the phase-screen technique.
At the same time, the single-time scenario also enables the formulation of analytical models and their verification against numerical data.

The existing analytical models can be classified into empirical ones and those based on explicit physical assumptions about the behavior of the beam shape and centroid.
Importantly, all parameters describing these models should, in principle, be derivable from the second- and fourth-order field-correlation functions.
The most critical parameters that analytical models should ideally reproduce are the first two moments of the transmittance.

Empirical models based on the log-normal and Beta distributions reproduce the first two transmittance moments by construction.
In this case, the Beta distribution generally provides a better fit to numerical data compared with the log-normal distribution.
Another class of models includes the beam-wandering, circular-beam, and elliptic-beam models.
These explicitly account for contributions from beam wandering and introduce assumptions about beam-spot distortion.
In all three cases, the beam profile at the receiver is assumed to have a Gaussian shape.
In the beam-wandering model, this profile does not fluctuate.
In the circular-beam model, the beam-spot radius fluctuates symmetrically, while in the elliptic-beam model the spot is modeled as a random ellipse.

All three models share a common shortcoming: the mode of their distribution is significantly biased compared with the numerical data.
This discrepancy arises from the strongly non-Gaussian instantaneous beam shape at the receiver.
The most natural way to improve the situation is inherent to the circular-beam model.
In this case, the model parameters can be matched to the first two moments of the transmittance.
The resulting model then explicitly depends on three parameters: the beam-wandering variance in addition to the two transmittance moments.
In principle, the same procedure could be applied to the beam-wandering model, but in that case both the beam-wandering variance and the squared beam-spot radius would be significantly biased.
For the elliptic-beam model, such an improvement appears impractical, since this model involves a more complicated analytical and numerical treatment.

The models based on the law of total probability explicitly account for the effect of beam wandering.
In addition, they empirically describe the contribution of beam-spot distortions using either a log-normal or a Beta distribution.
An advantage of these models is that they automatically reproduce the first two moments of the transmittance.
A drawback, however, is that they rely on strong analytical assumptions, which may lead to discrepancies with numerical data in certain scenarios.

The most serious drawback of applying analytical models is their high sensitivity to parameter errors.
Unfortunately, existing analytical methods often determine these parameters with significant inaccuracies.
This issue is particularly evident in the analytical expressions for the first two transmittance moments, which are valid only within a limited range of beam parameters and aperture sizes.
Although all the discussed analytical models are, in principle, applicable to any turbulence strength, here we restrict our attention to the case of weak turbulence.
This limitation highlights the need for more accurate analytical methods for evaluating the required parameters.
\acknowledgments % equivalent to \section*{ACKNOWLEDGMENTS}       
The authors thank D. Vasylyev, W. Vogel, M. Bohmann,  L. Bulla, P. Villoresi, A. Pilipenko, and B. Hage for enlightening discussions.
I.P. and A.A.S. acknowledge support from the National Academy of Sciences of Ukraine through Project 0125U000031 and from Simons Foundation International SFI-PD-Ukraine-00014573, PI LB.
M.K. acknowledges support from the National Research Foundation of Ukraine under Project No. 2023.03/0165.

%This unnumbered section is used to identify those who have aided the authors in understanding or accomplishing the work presented and to acknowledge sources of funding.  

% References
\bibliography{biblio} % bibliography data in report.bib

\begin{thebibliography}{10}

\bibitem{Brunner2014}
Brunner, N., Cavalcanti, D., Pironio, S., Scarani, V., and Wehner, S., ``Bell
  nonlocality,'' {\em Rev. Mod. Phys.}~{\bf 86},  419--478 (Apr 2014).

\bibitem{titulaer65}
Titulaer, U.~M. and Glauber, R.~J., ``Correlation functions for coherent
  fields,'' {\em Phys. Rev.}~{\bf 140},  B676--B682 (Nov. 1965).

\bibitem{mandel86}
Mandel, L., ``Non-classical states of the electromagnetic field,'' {\em Phys.
  Scr.}~{\bf 1986}(T12),  34 (1986).

\bibitem{mandel_book}
Mandel, L. and Wolf, E.,  [{\em Optical Coherence and Quantum
  Optics}{\nolinebreak\hspace{0.1em}]}, Cambridge University Press, Cambridge
  (1995).

\bibitem{vogel_book}
Vogel, W. and Welsch, D.-G.,  [{\em Quantum
  Optics}{\nolinebreak\hspace{0.1em}]}, Wiley-VCH Verlag GmbH \& Co. KGaA
  (2006).

\bibitem{agarwal_book}
Agarwal, G.~S.,  [{\em Quantum Optics}{\nolinebreak\hspace{0.1em}]}, Cambridge
  University Press, Cambridge (2013).

\bibitem{Schnabel2017}
Schnabel, R., ``Squeezed states of light and their applications in laser
  interferometers,'' {\em Phys. Rep.}~{\bf 684},  1--51 (2017).

\bibitem{sperling2018a}
Sperling, J. and Walmsley, I.~A., ``Quasiprobability representation of quantum
  coherence,'' {\em Phys. Rev. A}~{\bf 97},  062327 (Jun 2018).

\bibitem{sperling2018b}
Sperling, J. and Walmsley, I.~A., ``Quasistates and quasiprobabilities,'' {\em
  Phys. Rev. A}~{\bf 98},  042122 (Oct 2018).

\bibitem{sperling2020}
Sperling, J. and Vogel, W., ``Quasiprobability distributions for
  quantum-optical coherence and beyond,'' {\em Phys. Scr.}~{\bf 95},  034007
  (feb 2020).

\bibitem{gisin02}
Gisin, N., Ribordy, G., Tittel, W., and Zbinden, H., ``Quantum cryptography,''
  {\em Rev. Mod. Phys.}~{\bf 74},  145--195 (Mar 2002).

\bibitem{Xu2020}
Xu, F., Ma, X., Zhang, Q., Lo, H.-K., and Pan, J.-W., ``Secure quantum key
  distribution with realistic devices,'' {\em Rev. Mod. Phys.}~{\bf 92},
  025002 (May 2020).

\bibitem{Pirandola2020}
Pirandola, S., Andersen, U.~L., Banchi, L., Berta, M., D., B., Colbeck, R.,
  Englund, D., Gehring, T., Lupo, C., Ottaviani, C., Pereira, J.~L., Razavi,
  M., Shamsul~Shaari, J., Tomamichel, M., Usenko, V.~C., Vallone, G.,
  Villoresi, P., and Wallden, P., ``Advances in quantum cryptography,'' {\em
  Adv. Opt. Photon.}~{\bf 12},  1012--1236 (Dec 2020).

\bibitem{Renner2023}
Renner, R. and Wolf, R., ``Quantum advantage in cryptography,'' {\em AIAA
  Journal}~{\bf 61}(5),  1895--1910 (2023).

\bibitem{Tsang2016}
Tsang, M., Nair, R., and Lu, X.-M., ``Quantum theory of superresolution for two
  incoherent optical point sources,'' {\em Phys. Rev. X}~{\bf 6},  031033 (Aug
  2016).

\bibitem{Rouviere2024}
Rouvi\`{e}re, C., Barral, D., Grateau, A., Karuseichyk, I., Sorelli, G.,
  Walschaers, M., and Treps, N., ``Ultra-sensitive separation estimation of
  optical sources,'' {\em Optica}~{\bf 11},  166--170 (Feb 2024).

\bibitem{Semenov2024}
Semenov, A.~A., Samelin, J., Boldt, C., Sch\"unemann, M., Reiher, C., Vogel,
  W., and Hage, B., ``Photocounting measurements with dead time and afterpulses
  in the continuous-wave regime,'' {\em Phys. Rev. A}~{\bf 109},  013701 (Jan
  2024).

\bibitem{Uzunova2022}
Uzunova, V.~A. and Semenov, A.~A., ``Photocounting statistics of
  superconducting nanowire single-photon detectors,'' {\em Phys. Rev. A}~{\bf
  105},  063716 (Jun 2022).

\bibitem{ursin07}
Ursin, R. et~al., ``Entanglement-based quantum communication over 144 km,''
  {\em Nat. Phys.}~{\bf 3},  481 (Jul 2007).

\bibitem{fedrizzi09}
Fedrizzi, A., Ursin, R., Herbst, T., Nespoli, M., Prevedel, R., Scheidl, T.,
  Tiefenbacher, F., Jennewein, T., and Zeilinger, A., ``High-fidelity
  transmission of entanglement over a high-loss free-space channel,'' {\em Nat.
  Phys.}~{\bf 5},  389 (Jun 2009).

\bibitem{semenov10}
Semenov, A.~A. and Vogel, W., ``Entanglement transfer through the turbulent
  atmosphere,'' {\em Phys. Rev. A}~{\bf 81},  023835 (Feb. 2010).

\bibitem{gumberidze16}
Gumberidze, M.~O., Semenov, A.~A., Vasylyev, D., and Vogel, W., ``Bell
  nonlocality in the turbulent atmosphere,'' {\em Phys. Rev. A}~{\bf 94},
  053801 (Nov 2016).

\bibitem{perina73}
Peřina, J., Peřinov\'a, V., Teich, M.~C., and Diament, P., ``Two descriptions
  for the photocounting detection of radiation passed through a random medium:
  A comparison for the turbulent atmosphere,'' {\em Phys. Rev. A}~{\bf 7},
  1732--1737 (May 1973).

\bibitem{milonni04}
Milonni, P.~W., Carter, J.~H., Peterson, C.~G., and Hughes, R.~J., ``Effects of
  propagation through atmospheric turbulence on photon statistics,'' {\em J.
  Opt. B: Quantum Semiclass. Opt.}~{\bf 6}(8),  S742 (2004).

\bibitem{usenko12}
Usenko, V.~C., Heim, B., Peuntinger, C., Wittmann, C., Marquardt, C., Leuchs,
  G., and Filip, R., ``Entanglement of {G}aussian states and the applicability
  to quantum key distribution over fading channels,'' {\em New J. Phys.}~{\bf
  14}(9),  093048 (2012).

\bibitem{guo2017}
Guo, Y., Xie, C., Liao, Q., Zhao, W., Zeng, G., and Huang, D.,
  ``Entanglement-distillation attack on continuous-variable quantum key
  distribution in a turbulent atmospheric channel,'' {\em Phys. Rev. A}~{\bf
  96},  022320 (Aug 2017).

\bibitem{papanastasiou2018}
Papanastasiou, P., Weedbrook, C., and Pirandola, S., ``Continuous-variable
  quantum key distribution in uniform fast-fading channels,'' {\em Phys. Rev.
  A}~{\bf 97},  032311 (Mar 2018).

\bibitem{derkach2020b}
Derkach, I., Usenko, V.~C., and Filip, R., ``Squeezing-enhanced quantum key
  distribution over atmospheric channels,'' {\em New J. Phys.}~{\bf 22},
  053006 (may 2020).

\bibitem{hosseinidehaj2019}
Hosseinidehaj, N., Babar, Z., Malaney, R., Ng, S.~X., and Hanzo, L.,
  ``Satellite-based continuous-variable quantum communications:
  State-of-the-art and a predictive outlook,'' {\em IEEE Commun. Surv.
  Tutor.}~{\bf 21}(1),  881--919 (2019).

\bibitem{ShiyuWang2018}
Wang, S., Huang, P., Wang, T., and Zeng, G., ``Atmospheric effects on
  continuous-variable quantum key distribution,'' {\em New J. Phys.}~{\bf 20},
  083037 (aug 2018).

\bibitem{chai2019}
Chai, G., Cao, Z., Liu, W., Wang, S., Huang, P., and Zeng, G., ``Parameter
  estimation of atmospheric continuous-variable quantum key distribution,''
  {\em Phys. Rev. A}~{\bf 99},  032326 (Mar 2019).

\bibitem{Derkach2021}
Derkach, I. and Usenko, V.~C., ``Applicability of squeezed- and coherent-state
  continuous-variable quantum key distribution over satellite links,'' {\em
  Entropy}~{\bf 23}(1) (2021).

\bibitem{pirandola2021}
Pirandola, S., ``Composable security for continuous variable quantum key
  distribution: Trust levels and practical key rates in wired and wireless
  networks,'' {\em Phys. Rev. Research}~{\bf 3},  043014 (Oct 2021).

\bibitem{hosseinidehaj2021}
Hosseinidehaj, N., Walk, N., and Ralph, T.~C., ``Composable finite-size effects
  in free-space continuous-variable quantum-key-distribution systems,'' {\em
  Phys. Rev. A}~{\bf 103},  012605 (Jan 2021).

\bibitem{pirandola2021b}
Pirandola, S., ``Satellite quantum communications: Fundamental bounds and
  practical security,'' {\em Phys. Rev. Research}~{\bf 3},  023130 (May 2021).

\bibitem{pirandola2021c}
Pirandola, S., ``Limits and security of free-space quantum communications,''
  {\em Phys. Rev. Research}~{\bf 3},  013279 (Mar 2021).

\bibitem{peuntinger14}
Peuntinger, C., Heim, B., M\"uller, C.~R., Gabriel, C., Marquardt, C., and
  Leuchs, G., ``Distribution of squeezed states through an atmospheric
  channel,'' {\em Phys. Rev. Lett.}~{\bf 113},  060502 (Aug 2014).

\bibitem{Wang2018}
Wang, W., Xu, F., and Lo, H.-K., ``Prefixed-threshold real-time selection
  method in free-space quantum key distribution,'' {\em Phys. Rev. A}~{\bf 97},
   032337 (Mar 2018).

\bibitem{hofmann2019}
Hofmann, K., Semenov, A.~A., Vogel, W., and Bohmann, M., ``Quantum
  teleportation through atmospheric channels,'' {\em Phys. Scr.}~{\bf 94},
  125104 (sep 2019).

\bibitem{zhang2017}
Sheng-Li, Jin, C.-H., Shi, J.-H., Guo, J.-S., Zou, X.-B., and Guo, G.-C.,
  ``Continuous variable quantum teleportation in beam-wandering modeled
  atmosphere channel,'' {\em Chinese Phys. Lett.}~{\bf 34},  040302 (mar 2017).

\bibitem{villasenor2021}
Villase\~nor, E., He, M., Wang, Z., Malaney, R., and Win, M.~Z., ``Enhanced
  uplink quantum communication with satellites via downlink channels,'' {\em
  IEEE Trans. Quant. Eng.}~{\bf 2},  1--18 (2021).

\bibitem{bohmann16}
Bohmann, M., Semenov, A.~A., Sperling, J., and Vogel, W., ``Gaussian
  entanglement in the turbulent atmosphere,'' {\em Phys. Rev. A}~{\bf 94},
  010302(R) (Jul 2016).

\bibitem{hosseinidehaj15a}
Hosseinidehaj, N. and Malaney, R., ``Gaussian entanglement distribution via
  satellite,'' {\em Phys. Rev. A}~{\bf 91},  022304 (Feb 2015).

\bibitem{elser09}
Elser, D., Bartley, T., Heim, B., Wittmann, C., Sych, D., and Leuchs, G.,
  ``Feasibility of free space quantum key distribution with coherent
  polarization states,'' {\em New J. Phys.}~{\bf 11}(4),  045014 (2009).

\bibitem{heim10}
Heim, B., Elser, D., Bartley, T., Sabuncu, M., Wittmann, C., Sych, D.,
  Marquardt, C., and Leuchs, G., ``Atmospheric channel characteristics for
  quantum communication with continuous polarization variables,'' {\em Appl.
  Phys. B}~{\bf 98}(4),  635 (2010).

\bibitem{semenov12}
Semenov, A.~A., T\"oppel, F., Vasylyev, D.~Y., Gomonay, H.~V., and Vogel, W.,
  ``Homodyne detection for atmosphere channels,'' {\em Phys. Rev. A}~{\bf 85},
  013826 (Jan. 2012).

\bibitem{heim14}
Heim, B., Peuntinger, C., Killoran, N., Khan, I., Wittmann, C., Marquardt, C.,
  and Leuchs, G., ``Atmospheric continuous-variable quantum communication,''
  {\em New J. Phys.}~{\bf 16}(11),  113018 (2014).

\bibitem{glauber63c}
Glauber, R.~J., ``Coherent and incoherent states of the radiation field,'' {\em
  Phys. Rev.}~{\bf 131},  2766--2788 (Sep 1963).

\bibitem{sudarshan63}
Sudarshan, E. C.~G., ``Equivalence of semiclassical and quantum mechanical
  descriptions of statistical light beams,'' {\em Phys. Rev. Lett.}~{\bf 10},
  277--279 (Apr 1963).

\bibitem{semenov09}
Semenov, A.~A. and Vogel, W., ``Quantum light in the turbulent atmosphere,''
  {\em Phys. Rev. A}~{\bf 80},  021802(R) (Aug. 2009).

\bibitem{vasylyev12}
Vasylyev, D.~Y., Semenov, A.~A., and Vogel, W., ``Toward global quantum
  communication: Beam wandering preserves nonclassicality,'' {\em Phys. Rev.
  Lett.}~{\bf 108},  220501 (June 2012).

\bibitem{vasylyev16}
Vasylyev, D., Semenov, A.~A., and Vogel, W., ``Atmospheric quantum channels
  with weak and strong turbulence,'' {\em Phys. Rev. Lett.}~{\bf 117},  090501
  (Aug. 2016).

\bibitem{vasylyev18}
Vasylyev, D., Vogel, W., and Semenov, A.~A., ``Theory of atmospheric quantum
  channels based on the law of total probability,'' {\em Phys. Rev. A}~{\bf
  97},  063852 (June 2018).

\bibitem{klen2023}
Klen, M. and Semenov, A.~A., ``Numerical simulations of atmospheric quantum
  channels,'' {\em Phys. Rev. A}~{\bf 108},  033718 (Sept. 2023).

\bibitem{Klen2024}
Klen, M., Vasylyev, D., Vogel, W., and Semenov, A.~A., ``Time correlations in
  atmospheric quantum channels,'' {\em Phys. Rev. A}~{\bf 109},  033712 (Mar
  2024).

\bibitem{pechonkin2025}
Pechonkin, I., Klen, M., and Semenov, A.~A., ``Circular-beam approximation for
  quantum channels in the turbulent atmosphere,'' (2025).
\newblock arXiv:2507.12947 [quant-ph].

\bibitem{Charnotskii2013a}
Charnotskii, M., ``Sparse spectrum model for a turbulent phase,'' {\em J. Opt.
  Soc. Am. A}~{\bf 30},  479--488 (Mar 2013).

\bibitem{Charnotskii2013b}
Charnotskii, M., ``Statistics of the sparse spectrum turbulent phase,'' {\em J.
  Opt. Soc. Am. A}~{\bf 30},  2455--2465 (Dec 2013).

\bibitem{Charnotskii2020}
Charnotskii, M., ``Comparison of four techniques for turbulent phase screens
  simulation,'' {\em J. Opt. Soc. Am. A}~{\bf 37},  738--747 (May 2020).

\bibitem{Fleck1976}
Fleck, J.~A., Morris, J.~R., and Feit, M.~D., ``Time-dependent propagation of
  high energy laser beams through the atmosphere,'' {\em Appl. Phys.}~{\bf 10},
   129--160 (Jun 1976).

\bibitem{Frehlich2000}
Frehlich, R., ``Simulation of laser propagation in a turbulent atmosphere,''
  {\em Appl. Opt.}~{\bf 39},  393--397 (Jan 2000).

\bibitem{Lukin_book}
Lukin, V.~P. and Fortes, B.~V.,  [{\em Adaptive Beaming and Imaging in the
  Turbulent Atmosphere}{\nolinebreak\hspace{0.1em}]}, {SPIE} (Sept. 2002).

\bibitem{Schmidt_book}
Schmidt, J.,  [{\em Numerical simulations of optical wave propagation with
  examples in {MATLAB}}{\nolinebreak\hspace{0.1em}]}, SPIE, Bellingham (2010).

\bibitem{Tatarskii}
Tatarskii, V.,  [{\em The Effect of the Turbulent Atmosphere on Wave
  Propagation}{\nolinebreak\hspace{0.1em}]}, Israel Program for Scientific
  Translations, Jerusalem (1971).

\bibitem{Tatarskii2016}
Tatarskii, V.,  [{\em Wave propagation in a turbulent
  medium}{\nolinebreak\hspace{0.1em}]}, Dover Publications Inc., Mineola, New
  York (2016).

\bibitem{Fante1975}
Fante, R., ``Electromagnetic beam propagation in turbulent media,'' {\em Proc.
  IEEE}~{\bf 63}(12),  1669--1692 (1975).

\bibitem{Fante1980}
Fante, R., ``Electromagnetic beam propagation in turbulent media: An update,''
  {\em Proc. IEEE}~{\bf 68}(11),  1424--1443 (1980).

\bibitem{Andrews_book}
Andrews, L. and Phillips, R.,  [{\em Laser Beam Propagation Through Random
  Media}{\nolinebreak\hspace{0.1em}]}, Online access with subscription: SPIE
  Digital Library, Society of Photo Optical (2005).

\bibitem{taylor1938}
Taylor, G.~I., ``The spectrum of turbulence,'' {\em Proc. R. Soc. London
  A}~{\bf 164}(919),  476--490 (1938).

\bibitem{Aksenov1979}
Aksenov, V.~P. and Mironov, V.~L., ``Phase approximation of the
  huygens-kirchhoff method in problems of reflections of optical waves in the
  turbulent atmosphere,'' {\em J. Opt. Soc. Am.}~{\bf 69},  1609--1614 (Nov
  1979).

\bibitem{Banakh1979}
Banakh, V.~A. and Mironov, V.~L., ``Phase approximation of the
  huygens--kirchhoff method in problems of space-limited optical-beam
  propagation in turbulent atmosphere,'' {\em Opt. Lett.}~{\bf 4},  259--261
  (Aug 1979).

\bibitem{Mironov1977}
Mironov, V.~L. and Nosov, V.~V., ``On the theory of spatially limited light
  beam displacements in a randomlyinhomogeneous medium,'' {\em J. Opt. Soc.
  Am.}~{\bf 67},  1073--1080 (Aug 1977).

\bibitem{diament70}
Diament, P. and Teich, M.~C., ``Photodetection of low-level radiation through
  the turbulent atmosphere,'' {\em J. Opt. Soc. Am.}~{\bf 60},  1489--1494 (Nov
  1970).

\bibitem{perina72}
Pe{\v r}ina, J., ``On the photon counting statistics of light passing through
  an inhomogeneous random medium,'' {\em Czech. J. Phys. B}~{\bf 22}(11),
  1075--1084 (1972).

\bibitem{Berger2014}
Berger, V.~W. and Zhou, Y.,  [{\em {K}olmogorov--{S}mirnov {T}est:
  {O}verview}{\nolinebreak\hspace{0.1em}]}, John Wiley \& Sons, Ltd (2014).

\bibitem{Esposito1967}
Esposito, R., ``Power scintillations due to the wandering of the laser beam,''
  {\em Proc. IEEE}~{\bf 55}(8),  1533--1534 (1967).

\bibitem{Marcum_book}
Marcum, J.~I.,  [{\em Table of Q Functions}{\nolinebreak\hspace{0.1em}]}, RAND
  Corporation, Santa Monica, CA (1950).

\bibitem{Agrest_book}
Agrest, M. and Maximov, M.,  [{\em Theory of Incomplete Cylindrical Functions
  and their Applications}{\nolinebreak\hspace{0.1em}]}, Springer, Berlin
  (1971).

\bibitem{Vasylyev2013}
Vasylyev, D., Semenov, A.~A., and Vogel, W., ``Quantum channels with beam
  wandering: an analysis of the {M}arcum ${Q}$-function,'' {\em Phys.
  Scr.}~{\bf T153},  014062 (2013).

\bibitem{vasylyev17}
Vasylyev, D., Semenov, A.~A., Vogel, W., G\"unthner, K., Thurn, A., Bayraktar,
  O., and Marquardt, C., ``Free-space quantum links under diverse weather
  conditions,'' {\em Phys. Rev. A}~{\bf 96},  043856 (Oct 2017).

\end{thebibliography}
\bibliographystyle{spiebib} % makes bibtex use spiebib.bst

\end{document}